\documentclass{article}

\usepackage{microtype}
\usepackage{graphicx}
\usepackage{subcaption}
\usepackage{booktabs}
\usepackage{hyperref}
\usepackage{placeins}

\usepackage[accepted]{icml2026}

\usepackage{amsmath}
\usepackage{amssymb}
\usepackage{mathtools}

\usepackage{tabularx}
\usepackage{array}
\usepackage{ragged2e}
\newcolumntype{Y}{>{\RaggedRight\arraybackslash}X}
\icmltitlerunning{Context-Aware Server Collaboration for MCP}

\begin{document}

\twocolumn[
\icmltitle{Enhancing Model Context Protocol (MCP) with Context-Aware Server Collaboration}

\begin{icmlauthorlist}
    \icmlauthor{Meenakshi Amulya Jayanti}{uchicago}  
    \icmlauthor{X.Y. Han}{uchicago}                   
\end{icmlauthorlist}

\icmlaffiliation{uchicago}{University of Chicago}

\icmlcorrespondingauthor{Meenakshi Amulya Jayanti}{amulyajayanti@uchicago.edu}
\icmlcorrespondingauthor{X.Y. Han}{xy.han@chicagobooth.edu}   

\icmlkeywords{Model Context Protocol, Multi-Agent Systems, LLM Agents, Agent Coordination, Shared Context}

\vskip 0.3in
]

\printAffiliationsAndNotice{}

\begin{abstract}
The Model Context Protocol (MCP) \cite{mcp_docs} has emerged as a widely used framework for enabling LLM-based agents to communicate with external tools and services. The original MCP implementation \cite{anthropic_mcp} relies on a Large Language Model (LLM) to decompose tasks and issue instructions to servers. In particular, the agents, models, and servers are stateless and do not have access to a global context. However, in tasks involving LLM-driven coordination, it is natural that a {\it Shared Context Store (SCS)} could improve the efficiency and coherence of multi-agent workflows by reducing redundancy and enabling knowledge transfer between servers. Thus, in this work, we design and assess the performance of a {\it Context-Aware MCP (CA-MCP)} that offloads execution logic to specialized MCP servers that read from and write to a \textit{shared context memory}, allowing them to coordinate more autonomously in real time. In this design, context management serves as the central mechanism that maintains continuity across task executions by tracking intermediate states and shared variables, thereby enabling persistent collaboration among agents without repeated prompting. We present experiments showing that the CA-MCP can outperform the traditional MCP by reducing the number of LLM calls required for complex tasks and decreasing the frequency of response failures when task conditions are not satisfied. In particular, we conducted experiments on the {\it TravelPlanner} \cite{travelplanner} and {\it REALM-Bench} \cite{realm} benchmark datasets and observed statistically significant results indicating the potential advantages of incorporating a shared context store via CA-MCP in LLM-driven multi-agent systems.
\end{abstract}

\paragraph{Reproducibility.}
We provide a complete implementation of our proposed Context-Aware MCP system, including:  
(1)~planning using the central LLM,  
(2) Shared Context Store orchestration,  
(3) autonomous MCP server execution, and  
(4) evaluation pipelines used in Section~\ref{sec:evaluation}.  
All code and configurations are available in the anonymous GitHub repository \url{https://github.com/Research-Anonymous-25/Context-Aware-MCP-Framework} to enable independent replication of results.

\section{Introduction}
Traditional Large Language Model (LLM)-based multi-agent systems rely on an LLM acting as a central “planner” that directs specialized tool-using agents (MCP servers) through each step of a task~\cite{anthropic_mcp, mcp_docs, koul_mcp, openai_agents}. We will refer to this LLM as {\it Central LLM} of the MCP protocol and it effectively acts as an orchestrator. The Model Context Protocol (MCP), introduced by Anthropic~\cite{anthropic_mcp}, formalizes this orchestration model by specifying how a central LLM can communicate with external tool servers through structured JSON-based exchanges. In this design, the central LLM maintains the reasoning state, while the servers are stateless executors that perform specific actions and return results. This stateless separation simplifies tool integration but could also limit inter-agent awareness---preventing servers from leveraging shared context across tasks. As shown in Figure~\ref{fig:traditional}, the central LLM must operate within a fixed context window. This approach could incur significant computational overhead due to repeated inference calls for every subtask. Moreover, when used with a fixed context window, fully centralized control often leads all servers to send their entire contexts to the central planner---inducing context loss between steps and slower response times as the LLM struggles to coordinate all agents in dynamic environments. Recent works have identified these context management challenges as a key bottleneck in scaling AI agents to complex real-world tasks \cite{survey, ctxeng}. We hypothesize that these observed inefficiencies may originate from the fact that LLMs alone often forget constraints or details over a long context, causing cascading errors or hallucinations in multi-step planning. As tasks grow in complexity, a purely central planner-driven orchestration may become economically and operationally inefficient.

Motivated by these potential obstacles, in this work, we explore the potential of a \textbf{Context-Aware MCP (CA-MCP)}---an architectural modification of the traditional MCP, in which the central LLM is used only for high-level planning and final summarization, rather than constant step-by-step control. Our main proposal is introducing a \textbf{Shared Context Store (SCS)} accessible by all MCP servers and the central LLM. After the central LLM interprets the user's query and seeds this shared store with initial context (goals, parameters, etc.), the individual servers take over execution. These servers operate as collaborative agents that continuously read from and write to the shared context, coordinating with each other through this medium rather than through repeated prompts to the central LLM. The central LLM remains idle during the execution phase unless high-level guidance is required, and finally re-engages to read the accumulated context and produce the final response. Intuitively, decentralizing the orchestration in CA-MCP can overcome the scalability and latency issues of traditional MCP, while maintaining semantic accuracy and adherence to user constraints. More concretely, details of the architecture and implementation are provided in Sections~\ref{sec:architecture} and~\ref{sec:evaluation} with empirical results illustrated in Figures~\ref{fig:exec_time}--\ref{fig:rougel} and Tables~\ref{tab:travelplanner} and~\ref{tab:realm}.

\paragraph{Organization of the Paper.}
The remainder of this paper is structured as follows. In Section~\ref{sec:architecture}, we detail the CA-MCP architecture, contrasting it with the traditional MCP setup and highlighting the roles of the LLM and servers alongside the new shared context store. Section~\ref{sec:related} reviews related work, positioning our approach against other recent architectures such as RL-trained orchestrators~\cite{orchestrator}, SagaLLM~\cite{SagaLLM}, blackboard systems~\cite{blackboard}, Marvin memory~\cite{marvinmem}, and G-Memory~\cite{gmemory}. Section~\ref{sec:evaluation} presents an empirical evaluation of both the traditional and CA-MCP on (i) the TravelPlanner benchmark for itinerary planning and (ii) the Wedding Logistics problem from REALM-Bench, analyzing execution speed, solution quality, and coordination efficiency. Sections~\ref{sec:discussion} and~\ref{sec:conclusion} discuss the improvements achieved, system robustness, and future scalability, and conclude with directions for future work such as enabling parallelism, ensuring security, and integrating multimodal capabilities.

\paragraph{Limitations and Promising Future Work.}
In this paper, we focus specifically on evaluating whether introducing a shared context store into the Model Context Protocol (MCP) can reduce central LLM workload and improve coordination efficiency in multi-agent workflows. Our analysis centers on whether decentralizing orchestration—by shifting stepwise execution from the central LLM to collaborative servers—can mitigate latency, reduce repeated reasoning, and improve robustness in extended multi-step tasks. Another related hypothesis, which we do not fully explore here, is that many of the reported inefficiencies in traditional MCP could also be addressed through new architectural innovations, such as enhanced memory mechanisms \cite{legomem}, dynamic or multi-threaded coordination mechanisms, reinforcement-learning based orchestrators that learn task assignment and coordination policies \cite{metaorch}, and advanced multi-agent frameworks for role-based coordination and tool orchestration \cite{azure_agents_patterns}. For brevity, we limit our investigation to the context-aware modification of MCP, leaving these promising directions for future work.

\section{Detailed Architecture of CA-MCP}
\label{sec:architecture}

\begin{figure*}
\centering
\begin{subfigure}{0.8\linewidth}
    \centering
    \includegraphics[width=\linewidth]{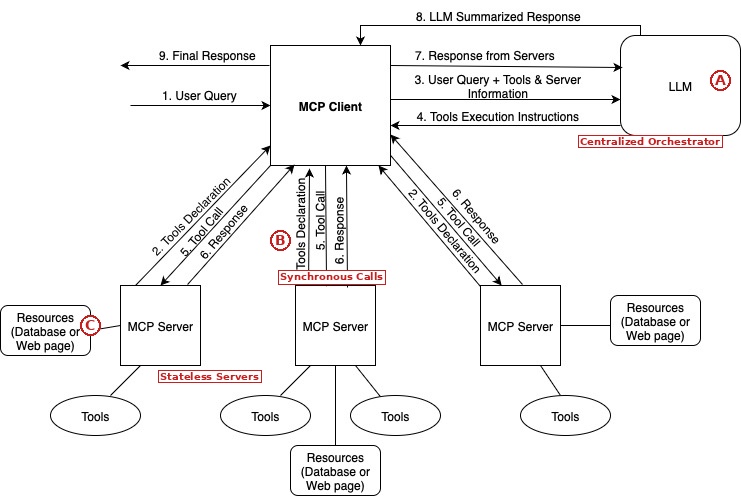}
    \caption{Traditional Model Context Protocol (MCP) architecture~\cite{anthropic_mcp}.
    The Central LLM acts as a centralized orchestrator (A), issuing step-by-step tool calls to stateless MCP servers (B) via synchronous interactions (C), tightly coupling execution to continuous LLM involvement.}
    \label{fig:traditional}
\end{subfigure}

\vspace{4pt}

\begin{subfigure}{0.8\linewidth}
    \centering
    \includegraphics[width=\linewidth]{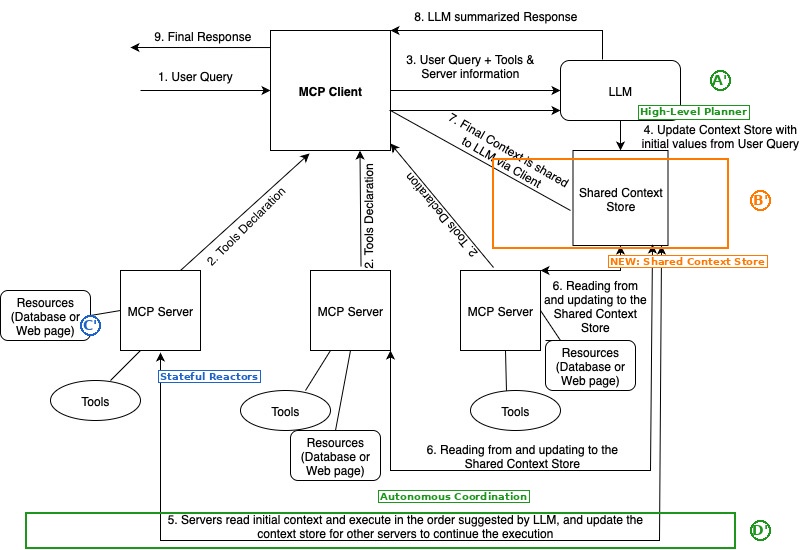}
    \caption{Our proposed CA-MCP architecture.
    A \emph{Shared Context Store (SCS)} (B') enables autonomous server coordination without continuous LLM involvement.
    The Central LLM functions mainly as planner and summarizer (A'), while servers operate as stateful reactors over shared state (C'), shifting execution from centralized orchestration to distributed coordination (D').}
    \label{fig:contextaware}
\end{subfigure}
\caption{Comparison of traditional MCP vs. our proposed CA-MCP.}
\end{figure*}

The CA-MCP architecture introduces a \textbf{Shared Context Store (SCS)} into the MCP workflow, fundamentally redefining how the central LLM and MCP servers interact. Unlike traditional MCP, where the central LLM is responsible for directing every action and maintaining full control of the workflow, the proposed design delegates operational autonomy to the servers while keeping the central LLM in the role of a strategic planner. Figures~\ref{fig:traditional} and~\ref{fig:contextaware} illustrate the key architectural differences between the two approaches.

\subsection{Key Components of CA-MCP}
Below, we describe the distinct roles of the different architectural components of CA-MCP and how this differs from the original MCP framework~\cite{anthropic_mcp}.
\paragraph{Central LLM (Long-Term Planner).}
The central Large Language Model assumes the role of the ``long-term planner'' of the system. Its initial engagement involves interpreting the user’s high-level goals and decomposing them into manageable tasks. During this phase, the central LLM selects the relevant MCP servers and seeds the SCS with an initial contextual blueprint (e.g., goals, constraints, execution outline). These actions correspond to Steps 1--4 in Figure~\ref{fig:contextaware}. Once this state is established, the central LLM reduces its involvement in fine-grained orchestration, allowing MCP servers to coordinate autonomously. The central LLM re-enters in Steps 7--9 to read the final state from the SCS and synthesize a complete response for the user.
{\it Comparison to Traditional MCP:} In the standard MCP setup~\cite{anthropic_mcp}, the central LLM stays involved throughout execution to issue each tool call and maintain task state. In CA-MCP, the central LLM performs high-level planning and final synthesis, while intermediate coordination is handled by MCP servers through the SCS.
\paragraph{MCP Servers (Short-Term Reactors).}
MCP servers behave as \emph{stateful reactors}. After tool declarations are made to the client (Step 2), each server monitors the SCS for relevant triggers needed to execute its functionality. Upon detecting suitable conditions, a server reads the current context, performs its computation or retrieval, updates the store with intermediate outputs, and signals readiness for subsequent tasks. These activities correspond to Steps 5--6 in Figure~\ref{fig:contextaware}. This event-driven self-coordination enables servers to adapt to changing conditions without repeated central LLM intervention.
{\it Comparison to Traditional MCP:} In the standard MCP framework~\cite{anthropic_mcp}, servers operate as stateless tools that respond only when invoked by the central LLM. In CA-MCP, servers actively participate by monitoring and updating the SCS.
\paragraph{Shared Context Store.}
The Shared Context Store (SCS) acts as a centralized blackboard~\cite{blackboard} for coordination, providing a single source of truth for task state, constraints, and intermediate outputs. The context store is initially seeded by the central LLM with the overall goals and structure of the workflow (Steps 3--4), after which MCP servers continuously read from and write to it during execution (Steps 5--6). Finally, the completed context is returned to the central LLM via the client for summarization and user response synthesis (Steps 7--9). By synchronizing through this store, MCP servers coordinate their actions without direct peer-to-peer messaging, making control emergent and distributed.
{\it Comparison to Traditional MCP:} The SCS does not appear in the traditional MCP. Figure~\ref{fig:contextaware} illustrates the resulting architecture, compared with the traditional system shown in Figure~\ref{fig:traditional}.

\subsection{Benefits and Conceptual Shift}
\begin{itemize}
    \item \textbf{Reduced central LLM Overhead:} Only two central LLM calls per workflow (plan and summarize), lowering latency and computational cost.
    \item \textbf{Improved Contextual Accuracy:} Persistent shared memory ensures constraints are preserved across the task sequence.
    \item \textbf{Robustness:} Distributed coordination avoids single points of failure and allows graceful handling of partial errors.
\end{itemize}

\FloatBarrier

\section{Related Work}
\label{sec:related}

LLM-based multi-agent systems employ various strategies for orchestration and inter-agent coordination, ranging from highly centralized to more distributed models. We review these paradigms and highlight how the CA-MCP both draws from and departs from them. 
\paragraph{Centralized Orchestration.} A prominent paradigm involves a centralized orchestrator \cite{orchestrator}, often referred to as a ``puppeteer,'' which dynamically directs agents based on the evolving task state. In this model, the central orchestrator maintains continuous control over agent actions and the global system state, dynamically routing agents at each step. The orchestrator's policy can even be updated using reinforcement learning (RL) to maximize efficiency and minimize redundancy, leading to more compact reasoning structures. A key distinction emerges when comparing this approach to our proposed CA-MCP. In RL-trained orchestrators, the central LLM actively and continuously routes agents throughout execution, implying persistent micro-management. While this flexibility allows adaptation to unexpected changes, it incurs significant computational cost due to continuous LLM inference calls. In contrast, our proposed CA-MCP limits the LLM’s role to initial planning and final summarization. By ``stepping back'' during execution and allowing servers to self-coordinate via the shared context, our design explicitly reduces redundant LLM calls while fostering autonomy among specialized agents.
\paragraph{Distributed and Hybrid Coordination.} Beyond purely centralized models, several frameworks enable agents to coordinate more autonomously. Surveys of LLM-based multi-agent systems (MAS) highlight structures such as ``shared message pools,'' which align conceptually with the SCS. These distributed or hybrid models \cite{survey} leverage collective intelligence, improving knowledge retention and enabling distributed handling of sub-tasks \cite{survey,llmcoord}. Our proposed CA-MCP represents a hybrid approach: centralized initialization by the LLM for coherence, followed by distributed execution through self-coordinating MCP servers. The SCS serves as a central hub for state exchange while decentralizing execution. This design balances the strengths of both paradigms—global planning from the LLM with scalable, event-driven adaptability from the servers. 
\paragraph{Blackboard Architecture.} The blackboard architecture \cite{blackboard}, a classic AI paradigm, presents a strong historical precedent. In this model, agents share information via a ``blackboard,'' a public memory space that replaces individual memory modules. Agents communicate exclusively through the blackboard, while a control unit (often an LLM agent) selects which agents act next. This directly addresses the burden of maintaining global context by externalizing shared state. However, unlike the blackboard system where an LLM controller remains continuously active, our proposed CA-MCP allows the LLM to disengage after seeding the context. Execution unfolds through self-coordination, avoiding persistent LLM involvement while retaining the benefits of a shared public memory. 
\paragraph{SagaLLM.} SagaLLM \cite{SagaLLM} introduces persistent memory to ensure transactional integrity, long-term consistency, and recovery from failures. Information such as goals, justifications, and compensation plans is stored in a structured ledger. A GlobalValidationAgent and SagaCoordinatorAgent manage state tracking and recovery, keeping orchestration heavily centralized. While both SagaLLM and CA-MCP involve shared state, their purposes diverge. SagaLLM emphasizes long-horizon workflows and robustness against failure through persistent, historical memory. Our proposed CA-MCP, by contrast, employs shared context as a real-time, transient workspace to enable reflexive, event-driven responses during execution. Its focus is efficiency and immediate adaptability rather than long-term consistency.
\paragraph{Marvin’s Memory.} Marvin \cite{marvin}, a Python framework for building LLM-based applications, provides persistent vector-based memory modules for agents. These allow agents to retain user preferences, build knowledge bases, and share conversational context. Marvin’s memory \cite{marvinmem} enriches individual agents but is not designed as a dynamic workspace for distributed execution. Our proposed CA-MCP instead emphasizes collective intelligence—servers coordinating through a common context store to reduce LLM orchestration overhead and improve execution efficiency. 
\paragraph{G-Memory.} G-Memory \cite{gmemory} introduces a hierarchical agentic memory, organizing history into Insight, Query, and Interaction Graphs. This enhances long-term learning, enabling agents to retrieve both high-level insights and detailed interaction trajectories across trials. While G-Memory focuses on self-evolution and cross-trial generalization, our proposed CA-MCP addresses a complementary problem: short-term coordination during a single workflow. Its goal is not learning across tasks but optimizing real-time execution by reducing redundant LLM mediation. 
\paragraph{Context Engineering.}
Context engineering has emerged as a practical discipline for structuring prompts, memory, tools, and state for LLM systems, with curated resources and design patterns documented in community efforts such as~\cite{ctxeng}. The SCS in our proposed CA-MCP can be seen as a concrete instantiation of these principles: by externalizing state, it achieves faster execution, reduced LLM calls, and greater scalability—directly addressing known limitations of prompt-only context handling.

\begin{table*}[t]
\caption{Comparison of CA-MCP with related paradigms in multi-agent orchestration and memory. 
Columns correspond to: RL-trained orchestrators~\cite{orchestrator},
SagaLLM~\cite{SagaLLM},
Blackboard Multi-Agent Systems~\cite{blackboard},
Marvin Memory~\cite{marvin},
and G-Memory~\cite{gmemory}.}
\label{tab:comparison}
\centering
\scriptsize
\renewcommand{\arraystretch}{1.5}
\begin{tabularx}{\textwidth}{|p{2.7cm}|Y|Y|Y|Y|Y|p{2.3cm}|}
\hline
\textbf{Feature} 
& \textbf{\textit{CA-MCP} (Our's)} 
& \textbf{RL-Trained Orchestrator} 
& \textbf{SagaLLM} 
& \textbf{Blackboard MAS} 
& \textbf{Marvin Memory} 
& \textbf{G-Memory} \\
\hline
Primary LLM Role 
& Initial planning + summarization only 
& Continuous dynamic orchestrator 
& Transaction coordinator with validation/rollback 
& Control unit selecting next agent 
& Enhancer of individual agents 
& Knowledge retriever across tasks \\
\hline
Agent Roles 
& Servers as autonomous reactors 
& Agents as puppets controlled by LLM 
& Domain agents under central coordinator 
& Contributors chosen by LLM 
& Agents with vector stores 
& Agents augmented by hierarchical graphs \\
\hline
Context Mechanism 
& Shared context store (dynamic, transient) 
& State aggregated by LLM 
& Persistent structured ledger 
& Blackboard (public memory) 
& Vector memory per agent 
& Query/Interaction graph \\
\hline
Coordination Style 
& Event-driven self-coordination 
& Continuous LLM routing 
& Centralized validation + rollback 
& LLM selects based on blackboard 
& Memory augmentation per agent 
& Cross-trial graph traversal \\
\hline
Focus 
& Efficiency, reduced LLM calls, real-time adaptation 
& Compact reasoning, flexible task routing 
& Long-term consistency, recovery 
& Token efficiency, modular agent reuse 
& Personalization and continuity 
& Long-term learning and generalization \\
\hline
\end{tabularx}
\end{table*}

\section{Empirical Evaluation}
\label{sec:evaluation}

To validate the advantages of our proposed CA-MCP, we evaluate its performance against the traditional MCP baseline on two benchmarks: (i) the \textbf{TravelPlanner \cite{travelplanner}} benchmark for real-world itinerary planning, and (ii) the \textbf{REALM-Bench \cite{realm}} Wedding Logistics (P5) scenario for multi-agent scheduling. We report metrics that capture execution efficiency, constraint satisfaction, semantic accuracy, and coordination effectiveness.

\subsection{Use Case 1: Travel Planning (TravelPlanner)}
\paragraph{Setup.}
The TravelPlanner \cite{travelplanner} benchmark provides over 1,200 multi-turn tasks that require generating realistic itineraries under budgetary, temporal, and preference constraints. We sample 500 queries such as: \emph{``Plan a three-day trip around Seattle with adventurous activities, vegan options, and a \$1500 budget.''} Each task requires integrating location recommendations, weather forecasts, hotel bookings, and dining suggestions.

\paragraph{Evaluation Metrics.}
We measure: \textbf{Execution Time} (wall-clock seconds), \textbf{Completeness} 
(0--1, fraction of pipeline stages completed), \textbf{BERTScore F1} 
\cite{bertscore} (semantic similarity), \textbf{ROUGE-L} \cite{rouge} (sequence 
overlap), and \textbf{LLM Calls} (orchestration overhead).

\paragraph{Traditional MCP.}
In the baseline, the LLM orchestrator centrally manages a set of stateless executors. Each server operates only when explicitly called, with no awareness of the broader workflow or other servers’ outputs. All coordination and state maintenance are handled by the LLM itself, which results in sequential execution, increased latency, and frequent loss of important context across long reasoning chains.

\paragraph{CA-MCP.}
In our approach, the central LLM is invoked once to parse constraints and seed the SCS with a structured JSON plan. MCP servers then self-orchestrate by reading triggers (e.g., \texttt{location\_done}) and updating the context with outputs. Once execution completes, the central LLM is called again for summarization.

\begin{figure*}[!t]
    \centering
    \includegraphics[width=\linewidth]{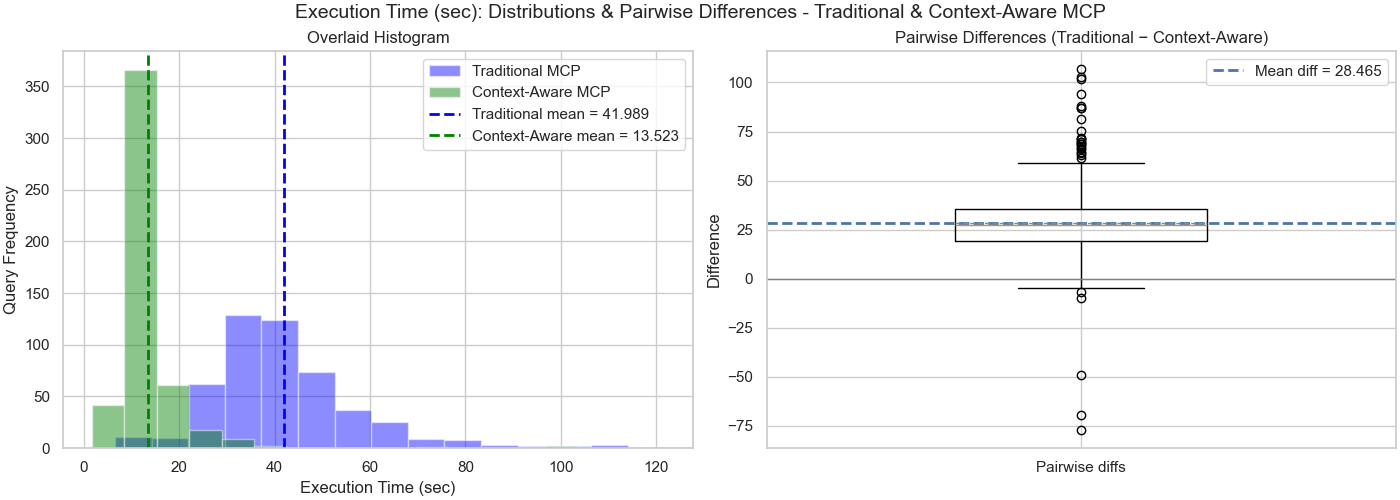}
    \caption{\textbf{Execution Time (sec)} on \textbf{TravelPlanner}~\cite{travelplanner} (500 queries). 
    (a) Overlaid histograms for \emph{Traditional} (blue) and \emph{Context-Aware} (green); dashed lines show means: 
    Traditional = 41.989, Context-Aware = 13.523. 
    (b) Pairwise differences \emph{(Traditional -- Context-Aware)}; positive values indicate Context-Aware is faster. 
    \textbf{Paired t-test:} \texttt{mean\_diff} = 28.465s, $\mathtt{sd} = 18.059 s$, $n = 500$, $\mathtt{pval} = 1.60e{-137}$. 
    \textbf{Takeaway:} CA-MCP markedly reduces latency.}
    \label{fig:exec_time}
\end{figure*}

\begin{figure*}[!t]
    \centering
    \includegraphics[width=\linewidth]{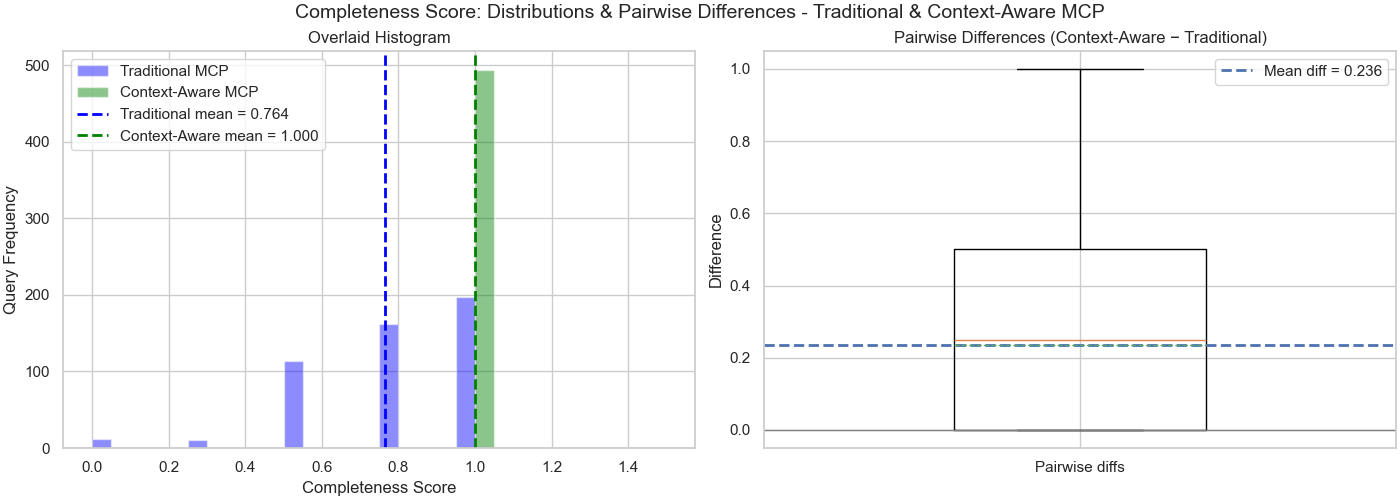}
    \caption{\textbf{Completeness Score} on \textbf{TravelPlanner}~\cite{travelplanner} (500 queries). 
    (a) Overlaid histograms for \emph{Traditional} (blue) and \emph{Context-Aware} (green); dashed lines show means:
    Traditional = 0.764, Context-Aware = 1.000. 
    (b) Pairwise differences \emph{(Context-Aware -- Traditional)}; positive values indicate higher completeness for Context-Aware. 
    \textbf{Paired t-test:} \texttt{mean\_diff} = 0.236, $\mathtt{sd} = 0.2408$, $n = 500$, $\mathtt{pval} = 4.39e{-75}$. 
    \textbf{Takeaway:} CA-MCP consistently attains higher completeness.}
    \label{fig:completeness}
\end{figure*}

\begin{figure*}[!t]
    \centering
    \includegraphics[width=\linewidth]{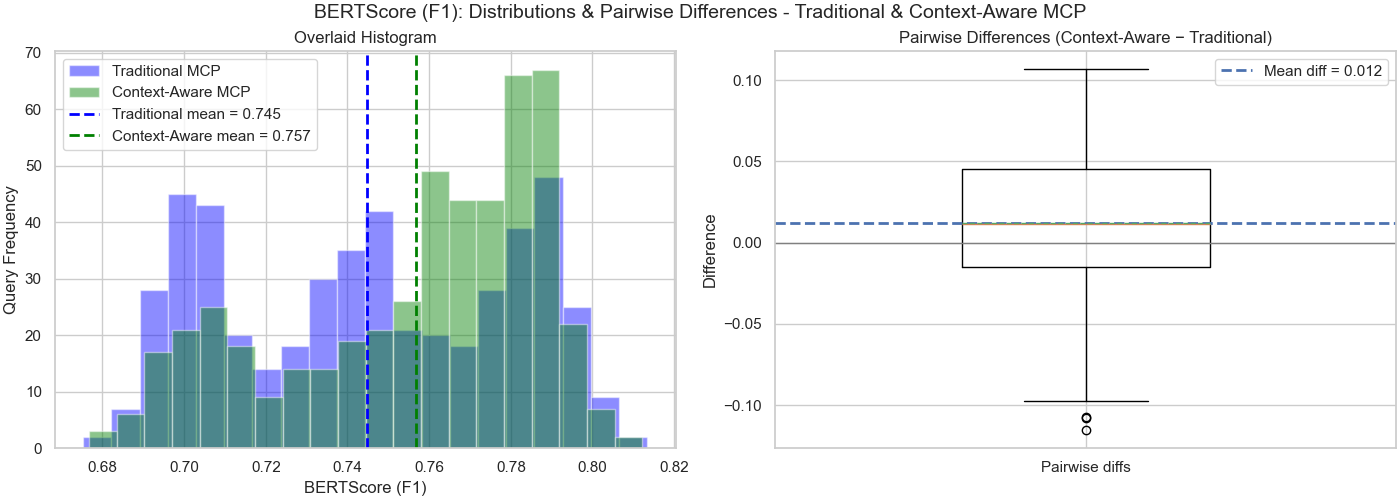}
    \caption{\textbf{BERTScore (F1)} on \textbf{TravelPlanner}~\cite{travelplanner} (500 queries). 
    (a) Overlaid histograms for \emph{Traditional} (blue) and \emph{Context-Aware} (green); dashed lines show means:
    Traditional = 0.745, Context-Aware = 0.757. 
    (b) Pairwise differences \emph{(Context-Aware -- Traditional)}.
    \textbf{Paired t-test:} \texttt{mean\_diff} = 0.012, $\mathtt{sd} = 0.04728$, $n = 500$, $\mathtt{pval} = 2.35e{-08}$. 
    \textbf{Takeaway:} CA-MCP yields a modest but consistent lift in semantic accuracy.}
    \label{fig:bertscore}
\end{figure*}

\begin{figure*}[!t]
    \centering
    \includegraphics[width=\linewidth]{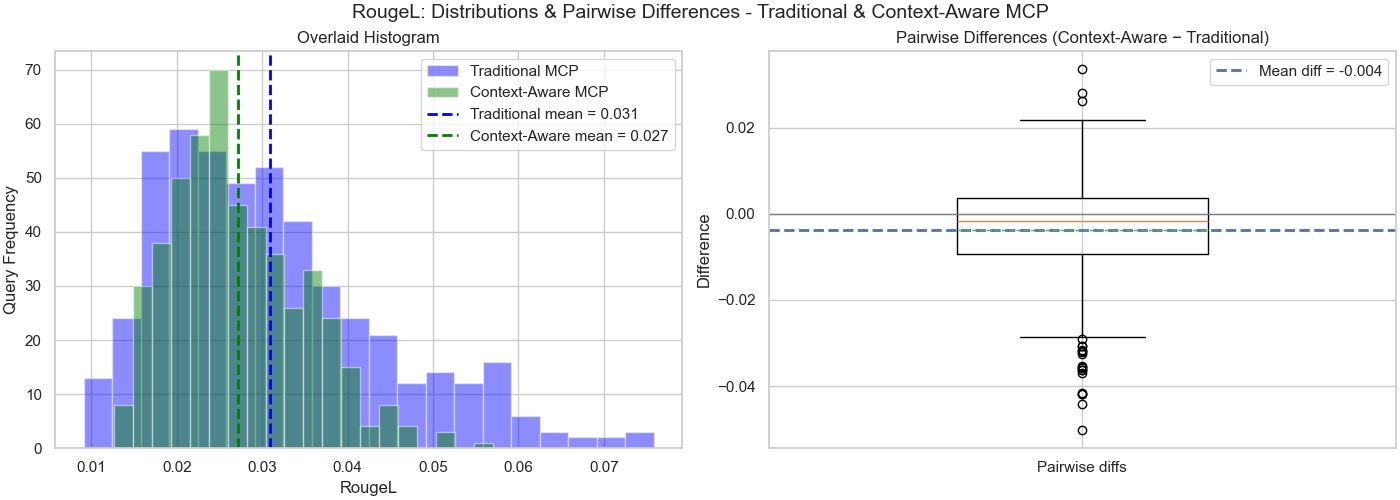}
    \caption{\textbf{ROUGE-L Score} on \textbf{TravelPlanner}~\cite{travelplanner} (500 queries). 
    (a) Overlaid histograms for \emph{Traditional} (blue) and \emph{Context-Aware} (green); dashed lines show means:
    Traditional = 0.031, Context-Aware = 0.027. 
    (b) Pairwise differences \emph{(Context-Aware -- Traditional)}. 
    \textbf{Paired t-test:} \texttt{mean\_diff} = $-0.004$, $\mathtt{sd} = 0.01173$, $n = 500$, $\mathtt{pval} = 1.24e{-13}$. 
    \textbf{Takeaway:} Differences are small, with a slight edge for Traditional.}
    \label{fig:rougel}
\end{figure*}

\paragraph{Results.}
As shown in Table~\ref{tab:travelplanner}, CA-MCP reduces average total execution time by 67.8\%, improves completeness of results, and slightly boosts semantic similarity scores while preserving output fluency.

\begin{table}[t]
\centering
\footnotesize
\setlength{\tabcolsep}{3pt}
\renewcommand{\arraystretch}{1.1}
\caption{\textbf{TravelPlanner}~\cite{travelplanner} results comparing Traditional and CA-MCP.}
\label{tab:travelplanner}
\begin{tabular}{lccl}
\toprule
\textbf{Metric} & \textbf{Traditional} & \textbf{CA-MCP} & \textbf{Improvement} \\
& \textbf{MCP} & & \\
\midrule
Execution Time (s) & 41.99 & 13.52 & \textbf{67.8\% faster} \\
Completeness (0--1) & 0.764 & 1.000 & \textbf{30.9\%} \\
BERTScore F1 & 0.745 & 0.757 & \textbf{0.012} \\
ROUGE-L Score & 0.031 & 0.027 & Comparable \\
LLM Calls & 5 & 2 & \textbf{60\% reduction} \\
\bottomrule
\end{tabular}
\end{table}

Thus, \textbf{Figure~\ref{fig:exec_time}} shows that a CA-MCP can provide significant speed advantages relative to a traditional MCP.

\subsection{Use Case 2: Wedding Logistics (REALM-Bench P5)}
\paragraph{Setup.}
REALM-Bench \cite{realm} is a rigorous benchmark for evaluating LLM-based multi-agent planning and scheduling in dynamic, real-world scenarios. We adapt problem P5 (Wedding Reunion), which requires orchestrating guest arrivals, errands, and shared vehicles under deadline and capacity constraints. The scenario challenges stateless LLM orchestration due to interdependencies and timing conflicts.

\paragraph{Evaluation Metrics.}
Following REALM-Bench \cite{realm}, we measure: \textbf{Goal Satisfaction} 
(0--1, objectives achieved), \textbf{Constraint Satisfaction} (0--1, hard 
constraints met), \textbf{Makespan} (minutes, schedule duration), 
\textbf{Coordination Score} (0/1, coordination effectiveness or batching indicator), \textbf{LLM Calls}, and \textbf{Execution Time} (seconds).

\paragraph{Traditional MCP.}
In the baseline, the central LLM orchestrator directly issues instructions to MCP servers. With no shared state between servers, all coordination flows through the LLM, which increases latency and loses opportunities for concurrency.

\paragraph{CA-MCP.}
Our implementation decomposes the problem into three MCP servers: \textsl{arrival\_tracker}, \textsl{errand\_tracker}, and \textsl{transport}. All communicate exclusively via the SCS, where transport requests and plans are dynamically updated. The central LLM initializes the shared state and later summarizes the final plan, but servers perform intermediate coordination independently.

\paragraph{Results.}
Table ~\ref{tab:realm} reports that both systems satisfy goals and constraints, but CA-MCP demonstrates substantially better efficiency and coordination, reducing makespan by 45.5\% and execution time by 73.5\%, while halving central LLM calls.

\begin{table}[t]
\centering
\footnotesize
\setlength{\tabcolsep}{3pt}
\renewcommand{\arraystretch}{1.1}
\caption{\textbf{REALM-Bench Wedding Logistics (P5)}~\cite{realm} results comparing Traditional and CA-MCP. CA-MCP achieves superior efficiency and coordination with fewer central LLM calls}
\label{tab:realm}
\begin{tabular}{lccl}
\toprule
\textbf{Metric} & \textbf{Traditional} & \textbf{CA-MCP} & \textbf{Improvement} \\
& \textbf{MCP} & & \\
\midrule
Goal Satisfaction & 1.0 & 1.0 & -- \\
Constraint Satisfaction & 1.0 & 1.0 & -- \\
Makespan (minutes) & 330.0 & 180.0 & \textbf{45.5\%} \\
Coordination Score (0/1) & 0 & 1 & \textbf{Improved} \\
LLM Calls & 2 & 1 & \textbf{50\%} \\
Execution Time (s) & 8.52 & 2.26 & \textbf{73.5\%} \\
\bottomrule
\end{tabular}
\end{table}

\section{Discussion}
\label{sec:discussion}

The experimental results across TravelPlanner and REALM-Bench highlight several important insights regarding the strengths and limitations of CA-MCP.

These results provide strong empirical evidence that CA-MCP improves execution efficiency while preserving solution quality:
\begin{itemize}
    \item \textbf{Efficiency:} Across both benchmarks, execution time dropped by 67--74\% due to reduced reliance on central LLM inference.
    \item \textbf{Optimality:} In REALM-Bench, makespan reduction and successful batching highlight superior scheduling efficiency.
    \item \textbf{LLM Call Reduction:} CA-MCP consistently required fewer central LLM invocations, mitigating context-window bottlenecks~\cite{llmcoord}.
    \item \textbf{Robustness:} Persistent shared context preserved user constraints, preventing cascading errors common in purely LLM-driven pipelines.
\end{itemize}

Together, these findings validate the central claim that introducing a shared context store into MCP workflows enhances scalability, robustness, and responsiveness in multi-agent execution.

\section{Future Work and Limitations}
\label{sec:futurework}

While the CA-MCP addresses critical pain points in current MCP design, several directions remain for future research:

\begin{itemize}
    \item \textbf{Parallel and Asynchronous Execution:} Enhancing the SCS to support asynchronous server execution, enabling large-scale parallel planning.
    \item \textbf{Server-Level Learning:} Allowing servers to adapt interaction strategies based on historical execution traces, further reducing dependence on the central LLM.
    \item \textbf{Multimodal Integration:} Extending the framework to include non-textual servers (e.g., vision, audio, structured data) contributing to shared context.
\end{itemize}

By pursuing these directions, CA-MCP can evolve into a robust, production-ready architecture for next-generation LLM-based multi-agent systems.

\paragraph{Limitations.} In our experiments (Tables \ref{tab:comparison}--\ref{tab:realm} and Figures \ref{fig:exec_time}--\ref{fig:rougel}), we used straightforward realizations of Traditional MCP (Figure~\ref{fig:traditional}) and CA-MCP (Figure~\ref{fig:contextaware}) to isolate the specific benefit of the SCS without hardware or software optimizations such as parallel multithreading. These optimizations could improve performance of both approaches and change their relative ordering. We leave such engineering explorations to future work. The goal of this study is to demonstrate that the conceptual idea of CA-MCP improves efficiency over a straightforward implementation of Traditional MCP.

\section{Conclusion}
\label{sec:conclusion}

This paper introduced CA-MCP, an extension of the Model Context Protocol that 
reallocates roles between a central LLM and specialized MCP servers. Our SCS enables autonomous server coordination, reducing redundant LLM calls and improving execution efficiency.

Empirical evaluation on TravelPlanner and REALM-Bench suggests that CA-MCP shows substantial improvements: 
up to 73.5\% faster execution, perfect completeness scores, and 45.5\% shorter 
makespan. These results suggest that shifting orchestration from the central LLM to 
shared state can be promising, though broader assessment is needed to fully characterize 
performance and generality.

Together, these results establish CA-MCP as a practical and scalable design for 
LLM-driven multi-agent systems, bridging centralized orchestration and distributed 
autonomy.

\section*{Impact Statement}

This paper advances multi-agent LLM orchestration through architectural improvements 
reducing computational overhead by 67.8\% (execution time) and 60\% (LLM calls). 
These gains could reduce deployment costs and environmental impact of AI systems. 
As with any AI advancement, improved efficiency could enable both beneficial and 
harmful applications; deployment should include appropriate safeguards and 
domain-specific ethical review. This foundational research does not introduce 
novel ethical concerns beyond those inherent to advancing AI agent capabilities.

\bibliographystyle{icml2026}
\bibliography{references}

@misc{anthropic_mcp,
  author = {{Anthropic}},
  title  = {Introducing the {Model Context Protocol}},
  year   = {2024},
  url    = {https://www.anthropic.com/index/model-context-protocol},
  note   = {Accessed 2025}
}

@misc{mcp_docs,
  author = {{MCP Community}},
  title  = {{Model Context Protocol}: Official Documentation and Specification},
  year   = {2025},
  url    = {https://www.modelcontextprotocol.io},
  note   = {Accessed 2025}
}

@misc{koul_mcp,
  author       = {Koul, Nimrita},
  title        = {The {Model Context Protocol} ({MCP}): A Complete Tutorial},
  year         = {2025},
  howpublished = {Medium},
  url          = {https://medium.com/@nimritakoul/mcp-complete-tutorial}
}

@misc{openai_agents,
  author = {{OpenAI}},
  title  = {{OpenAI} Agents {SDK} Documentation},
  year   = {2025},
  url    = {https://openai.github.io/openai-agents/mcp.html}
}

@article{survey,
  author  = {Tran, K.-T. and Dao, D. and Nguyen, M.-D. and Pham, Q.-V. and O'Sullivan, B. and Nguyen, H. D.},
  title   = {Multi-Agent Collaboration Mechanisms: A Survey of {LLM}s},
  journal = {arXiv preprint arXiv:2501.06322},
  year    = {2025}
}

@misc{ctxeng,
  author       = {Meirtz},
  title        = {Awesome Context Engineering},
  howpublished = {GitHub repository},
  year         = {2025},
  url          = {https://github.com/meirtz/awesome-context-engineering},
  note         = {Accessed Aug 5, 2025}
}

@article{blackboard,
  author  = {Han, B. and Zhang, S.},
  title   = {Exploring Advanced {LLM} Multi-Agent Systems Based on Blackboard Architecture},
  journal = {arXiv preprint arXiv:2507.01701},
  year    = {2025}
}

@inproceedings{bertscore,
  author    = {Zhang, Tianyi and Kishore, Varsha and Wu, Felix and Weinberger, Kilian Q. and Artzi, Yoav},
  title     = {{BERT}Score: Evaluating Text Generation with {BERT}},
  booktitle = {International Conference on Learning Representations},
  year      = {2020},
  url       = {https://openreview.net/forum?id=SkeHuCVFDr}
}

@article{orchestrator,
  author  = {Dang, Y. and Qian, C. and Luo, X. and others},
  title   = {Multi-Agent Collaboration via Evolving Orchestration},
  journal = {arXiv preprint arXiv:2505.19591},
  year    = {2025}
}

@article{SagaLLM,
  author  = {Chang, E. Y. and Geng, L.},
  title   = {{SagaLLM}: Context Management, Validation, and Transaction Guarantees for Multi-Agent {LLM} Planning},
  journal = {arXiv preprint arXiv:2503.11951},
  year    = {2025}
}

@misc{marvin,
  author = {{Marvin AI}},
  title  = {Marvin {AI} Framework},
  year   = {2025},
  url    = {https://askmarvin.ai/},
  note   = {Accessed Aug 5, 2025}
}

@misc{marvinmem,
  author = {{Marvin AI}},
  title  = {Marvin Documentation: Memory Pattern},
  year   = {2025},
  url    = {https://askmarvin.ai/patterns/memory},
  note   = {Accessed Aug 5, 2025}
}

@article{gmemory,
  author  = {Zhang, G. and Fu, M. and Wan, G. and others},
  title   = {{G-Memory}: Tracing Hierarchical Memory for Multi-Agent Systems},
  journal = {arXiv preprint arXiv:2506.07398},
  year    = {2025}
}

@inproceedings{llmcoord,
  author    = {Agashe, S. and Fan, Y. and Reyna, A. and Wang, X. E.},
  title     = {{LLM}-Coordination: Evaluating and Analyzing Multi-agent Coordination Abilities in Large Language Models},
  booktitle = {Findings of {NAACL}},
  year      = {2025}
}

@article{travelplanner,
  author  = {Yang, F. and Chen, Z. and Jiang, Z. and others},
  title   = {{TravelPlanner}: A Benchmark for Real-World Planning with Language Agents},
  journal = {arXiv preprint arXiv:2402.01622},
  year    = {2024}
}

@article{realm,
  author  = {Geng, L. and Chang, E. Y.},
  title   = {{REALM}-Bench: A Benchmark for Evaluating Multi-Agent Systems on Real-World, Dynamic Planning and Scheduling Tasks},
  journal = {arXiv preprint arXiv:2502.18836},
  year    = {2025}
}

@inproceedings{rouge,
  author    = {Lin, Chin-Yew},
  title     = {{ROUGE}: A Package for Automatic Evaluation of Summaries},
  booktitle = {Text Summarization Branches Out},
  pages     = {74--81},
  year      = {2004},
  address   = {Barcelona, Spain},
  publisher = {Association for Computational Linguistics}
}

@article{legomem,
  author  = {Han, D. and Couturier, C. and Madrigal Diaz, D. and Zhang, X. and R{\"u}hle, V. and Rajmohan, S.},
  title   = {{LEGOMem}: Modular Procedural Memory for Multi-agent {LLM} Systems for Workflow Automation},
  journal = {arXiv preprint arXiv:2510.04851},
  year    = {2025}
}

@article{metaorch,
  author  = {Agrawal, K. and Nargund, N.},
  title   = {Neural Orchestration for Multi-Agent Systems: A Deep Learning Framework for Optimal Agent Selection in Multi-Domain Task Environments},
  journal = {arXiv preprint arXiv:2505.02861},
  year    = {2025}
}

@misc{azure_agents_patterns,
  author       = {{Microsoft Azure Architecture Center}},
  title        = {{AI} Agent Orchestration Patterns},
  howpublished = {Microsoft Learn Documentation},
  year         = {2025},
  url          = {https://learn.microsoft.com/en-us/azure/architecture/ai-ml/guide/ai-agent-design-patterns},
  note         = {Accessed Nov 2025}
}

\end{document}